\def\etal{et~al.~}
\def\Lya{{\rm Ly}\kern 0.1em$\alpha$}
\def\Mg{{\rm Mg}\kern 0.1em{\sc ii}}
\def\Mgo{{\rm Mg}\kern 0.1em{\sc i}}
\def\Fe{{\rm Fe}\kern 0.1em{\sc ii}}
\def\MgII{{\rm Mg}\kern 0.1em{\sc ii}~$\lambda\lambda2976, 2803$}
\def\C{{\rm C}\kern 0.1em{\sc iv}}
\def\CIV{C\kern 0.1em{\sc iv}~$\lambda\lambda1548, 1550$}
\def\HI{{\rm H}\kern 0.1em{\sc i}}
\def\kms{\hbox{km~s$^{-1}$}}
\def\cm2{\hbox{cm$^{-2}$}}
\documentstyle[11pt,aaspp4,flushrt]{article}
\tighten

\slugcomment{Accepted: {\it Astrophysical Journal}}
\lefthead{Churchill, Steidel, \& Vogt}
\righthead{Mg II Absorbing Galactic Gas}

\begin{document}

\title{{\Large\bf On the Spatial and Kinematic Distributions of \\ 
        Mg~II Absorbing Gas in $\left< z \right> \sim 0.7$ 
        Galaxies}\footnote{Based upon observations obtained at the 
        W. M. Keck Observatory, which is jointly operated by the 
        University of California and the California Institute
        of Technology.}}
\thispagestyle{empty}

\author{{\sc Christopher W. Churchill}\altaffilmark{2,3},           
{\sc Charles C. Steidel}\altaffilmark{4,5,6}, {\sc and}                   
{\sc Steven S. Vogt}\altaffilmark{3,7}}                             
                                                              
\altaffiltext{2}{Department of Astronomy and Astrophysics,
                 Pennsylvania State University}               
\altaffiltext{3}{Board of Studies in Astronomy and Astrophysics,
       University of California, Santa Cruz}                              
\altaffiltext{4}{Palomar Observatory, Mail Stop 105--24,      
       California Institute of Technology}                    
\altaffiltext{5}{Alfred P. Sloan Foundation Fellow}           
\altaffiltext{6}{NSF Young Investigator}                      
\altaffiltext{7}{UCO/Lick Observatories}              

\begin{abstract}
We present HIRES/Keck spectra having resolution $\sim 6$~{\kms} of
{\Mg} $\lambda 2796$ absorption profiles which arise in the gas
believed to be associated with 15 identified galaxies over the
redshift range $(0.5 \leq z \leq 0.9)$.  
These galaxies have measured redshifts consistent with those seen in
absorption.
Using non--parametric rank correlation tests, we searched for
correlations of the absorption strengths, saturation, and
line--of--sight kinematics with the galaxy redshifts, rest frame $B$
and $K$ luminosities, rest $\left< B-K\right>$ colors, and impact
parameters $D$.
We found no correlations at the 2.5$\sigma$ level between the measured
absorption properties and galaxy properties.
Of primary significance is the fact that the QSO--galaxy impact
parameter apparently does not provide the primary distinguishing
factor by which absorption properties can be characterized.
The absorption properties of {\Mg} selected galaxies exhibit a large
scatter, which, we argue, is suggestive of a picture in which the gas
in galaxies arises from a variety of on--going dynamical events.
Inferences from our study include: 
(1) The spatial distribution of absorbing gas in and around galaxies
does not appear to follow a simple galactocentric functional
dependence, since the gas distribution is probably highly structured.  
(2) A {\it single} systematic kinematic model apparently cannot
describe the observed velocity spreads in the absorbing gas.
It is more likely that galaxy/halo events giving rise to absorbing gas
each exhibit their own systematic kinematics, so that a heterogeneous
population of sub--galaxy scale structures are giving rise to the 
observed cloud velocities.
(3) The absorbing gas spatial distribution and overall kinematics 
may depend upon gas producing events and mechanisms that are
recent to the epoch at which the absorption is observed.
In any given galaxy, these distributions likely change over a $\sim$~few
Gyr timescale (few dynamical times of the absorbing clouds), which 
provides one source for the observed scatter in the absorption
properties.
Based upon these inferences, we note that any evolution in the absorption
gas properties over the wider redshift range ($0.4\leq z\leq 2.2$)
should be directly quantifiable from a larger dataset of
high--resolution absorption profiles.
\end{abstract}

\keywords{galaxies: kinematics and dynamics --- 
          galaxies: evolution ---
          quasars: absorption lines}

\section{Introduction}
\pagestyle{myheadings}                                            
\markboth{\sc \hfill Churchill, Steidel, \& Vogt \hfill}
         {\sc \hfill Churchill, Steidel, \& Vogt \hfill}

The {\MgII} doublet, as seen in absorption in the spectra of background
QSOs (cf.~\cite{ss92}), is known to arise in the low ionization gas
associated with a population of ``normal'' field galaxies
that exhibit little to no evolution in their rest frame $\left<
B-K\right>$ colors since a redshift of $z \sim 1.0$ (\cite{sdp94},
hereafter SDP).
Galaxies selected by the known presence of {\Mg} absorption represent
a wide range of colors, from late--type spirals to the reddest
ellipticals, have $L_B$ and $L_K$ luminosity functions consistent with 
local luminosity functions in which morphological types later then
Sd are excluded, and exhibit a strong correlation between color and $L_K$ 
(fainter galaxies are bluer).
These facts suggests that a wide range of morphological types are
undergoing the processes that give rise to {\Mg} absorbing gas, but
that isolated low mass ($L_K < 0.07L_K^{\ast}$) ``faint blue
galaxies'' are not.

Now that we have a first look at what types of galaxies are being
selected by their {\Mg} absorption cross section, we can undertake
studies from which we hope to infer the spatial distribution and
line--of--sight kinematics of the absorbing gas and to examine
connections, if any, between the absorbing gas and galaxy properties.
If correlations between galaxy and absorption properties can be
established, where galaxies are directly accessible to imaging and
spectroscopy, then we may be able to infer the properties of higher
redshift galaxies simply by classifying their absorption properties.
The hope is that we may chart the general evolution of galaxies and
of their gas kinematic, chemical, and ionization conditions back to
the epoch of the earliest QSOs (\cite{steidel93a}; \cite{bergeron94};
\cite{welty96}; \cite{charlton96}; \cite{bechtold96}; \cite{halopaper}).

A more direct motive for this study is to examine the model of $z \leq
1$ galaxy halos suggested by Lanzetta \& Bowen (1990, 1992).
They suggested that intermediate redshift galactic halos are roughly
identical, have absorbing gas spatial number density
distributions $\propto r^{1-2}_{\rm gal}$, and likely exhibit
systematic rotational or radial flow kinematics.
The primary test for this picture is the prediction that the {\it
observed}\/ differences in the absorption properties from one system
to another are predominantly due to the QSO--galaxy impact parameter.

Based upon several studies (\cite{pb90}; \cite{bb91};
\cite{lanzetta92}; \cite{lebrun93}; \cite{steidel93b}; \cite{bbp95};
\cite{s95}), a consensus has emerged in which the ``galaxy/halo''
model (first proposed by \cite{bachall69}) is favored over the
``galactic fragments'' (cf.~\cite{yanny92}) and ``dwarf satellites''
(\cite{york86}) models.
Whatever the true nature of {\Mg} absorbing gas, the bottom line is
that a relatively luminous galaxy within $\sim 40$~kpc of the QSO line
of sight appears to be a prerequisite for the detection of {\Mg}
absorption (\cite{s95}).
If all scenarios contribute at some level, statistical correlations
between absorption and galaxy properties predicted by the galaxy/halo
model (\cite{lb92}; \cite{lanzetta92}; \cite{cc96}) would be
``diluted'' by the more stochastic kinematics expected from merging
dwarf satellite galaxies or accreting {\it gaseous}\/ sub--galactic
fragments. 
That is, if the local environments around {\Mg} absorbing galaxies are
populated by satellite galaxies and by fragments of on--going galaxy/halo
events (i.e.~accretion, superbubbles, merging), systematic trends with
impact parameter, or any other galaxy property, would be difficult to
detect due to a large scatter in the absorption properties.

In this paper, we report on observations of the {\Mg} ($\lambda 2796$)
absorption line obtained with a resolution of $\sim 6$~{\kms} and 
signal--to--noise ratios $\sim 30$, and explore correlations
between the absorption properties and the IR/optical luminosities
and colors, impact parameters, and redshifts of the identified
associated galaxies.

\section{Observations}

We observed the 11 QSOs presented in Table 1 with the HIRES echelle
spectrometer (\cite{vogt94}) on the Keck 10--m telescope on 1995
January 23--25. 
The spectral resolution was 6.6~{\kms} ($R=45,000$).  
There are small gaps in the spectra redward of 5100~{\AA} where the
free spectral range of the echelle format exceeds the width of the 
$2048 \times 2048$ Tektronix CCD.
The spectroscopic sample was taken from the absorption line survey of
Steidel \& Sargent (1992) and was selected on the basis that each
absorption system was associated with an imaged galaxy that had been 
spectroscopically confirmed to have the same redshift as seen in
absorption (SDP).

The galaxy properties were obtained from broad--band $g (4900/700)$,
$\Re (6930/1500)$, $i(8000/1450)$ and NICMOS images of the QSO fields.
This combination of filters allowed the rest--frame $B$ and $K$
magnitudes of the galaxies to be determined.
A brief description of the imaging and follow--up spectroscopic work
has been presented by SDP and Steidel \& Dickinson (1995).
Here, we concentrate on the properties of the 15 identified $z \leq 1$
absorbing galaxies for which we have acquired HIRES QSO absorption
line spectra.

\section{Analysis}

The HIRES data were reduced with the IRAF\footnote{IRAF is distributed by
the National Optical Astronomy Observatories, which are operated by
AURA, Inc., under contract to the NSF.} Apextract package for echelle
data.
We have identified absorption features in the spectra using the method
presented by Lanzetta, Turnshek, \& Wolfe (1987), where we have chosen
a bandpass of 2.3 resolution elements (i.e.~7 pixels).
A detection level of 5$\sigma$ was enforced and features having {\Mg}
rest equivalent widths $W_{\rm min} < 0.023$~{\AA} were discarded.
This latter criterion enforced a uniform detection level upon all
absorption systems at the level of the lowest signal--to--noise ratio
spectrum in the sample.

\begin{figure}[bh]                                                
\plotone{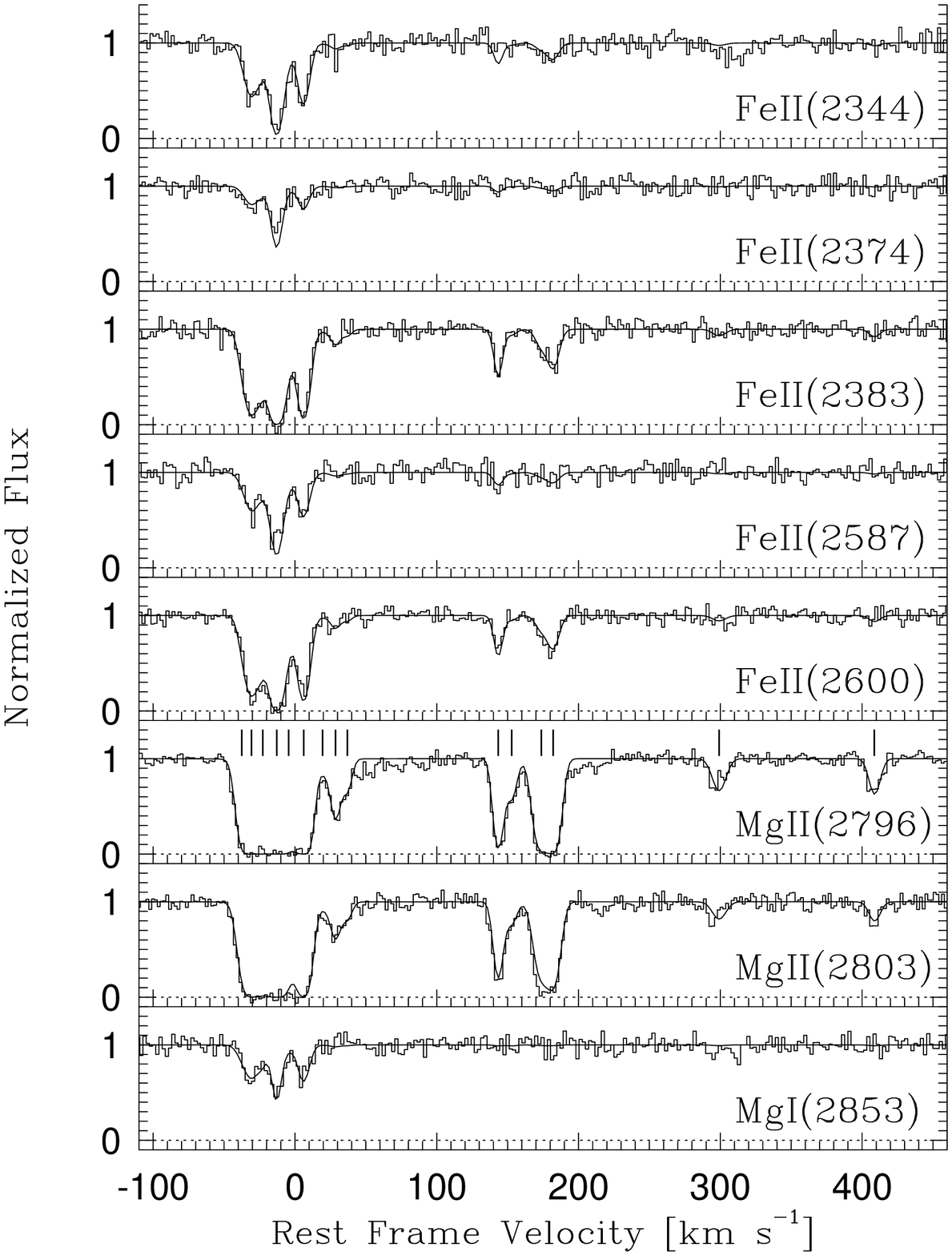}                                                  
\caption                                                          
{The multiple transitions observed for G2 with simultaneous       
profile fits shown. The subcomponents from the fit, and their     
velocities, are shown as vertical ticks above the {\Mg} (2796)    
transition.}                                                      
\end{figure}                                                      

Voigt profiles (natural + thermal broadening) convolved with the
instrumental broadening (Gaussian with $\sigma = 2.46$~{\kms}) were
fit to the profile substructures using an interactive profile--fitting
algorithm of our own design (\cite{mythesis}).
In all cases, the {\MgII} doublet and at least one unsaturated weaker
transition from either {\Fe} or {\Mgo} were fit simultaneously.
Thus, in the cases where the {\Mg} profiles appear saturated, there was
little ambiguity in both the number of subcomponents and their
relative velocities.
In Figure 1, we present the profile fits to the QSO absorption lines
observed from G2, which is the $z=0.8514$ galaxy in the Q0002+051
field.
Of the 15 systems used for this study, these profiles are the most
complex.
As such, this system serves as an illustration of the level of
information available for our profile fits and the ``accuracy'' of the
these fits over a range of subcomponent structures.

In Figure 2,\notetoeditor{We would like Figure 2 to be landscape full
page.} we present the HIRES {\Mg} ($\lambda 2796$) absorption
profiles in units of rest frame line--of--sight velocity.
The velocity spread of each panel is 400~{\kms}, except for G2, which
is 800~{\kms}.  
For illustration purposes (to allow a comparison of absorption
strengths and subcomponent variations with the projected galactocentric
distance probed by the line of sight), we have sequenced the profiles
in order of increasing QSO--galaxy impact parameter,
$Dh^{-1}$~kpc\footnote{We assume $q_0 = 0.05$ and quote all distances
in terms of $h = H_0/100$~{\kms}~Mpc$^{-1}$.}.

To quantitatively ascertain if any connections between the low
ionization gas absorption strengths and line--of--sight kinematics and
the galaxy properties are present in our sample, we characterized the
profiles by the number of subcomponents, $N_{\rm c}$, or ``clouds'',
and three simple kinematic indicators.
Since we assume no {\it a priori}\/ knowledge of the gas kinematics,
we have developed crude empirical indicators that are model
independent.
The first kinematic indicator is the ``absolute deviation from the
median'' of the cloud velocities,
\begin{equation}
A({\Delta v}) = \frac{1}{N_{\rm c}-1} 
                 \sum_{i=1}^{N_{\rm c}} \left| \Delta v_{i} \right| ,
\end{equation}
where $\Delta v_{i} = v_{i}-\bar{v}$, and $\bar{v}$ is the median cloud
velocity.

\begin{figure}[bh]                                              
\plotone{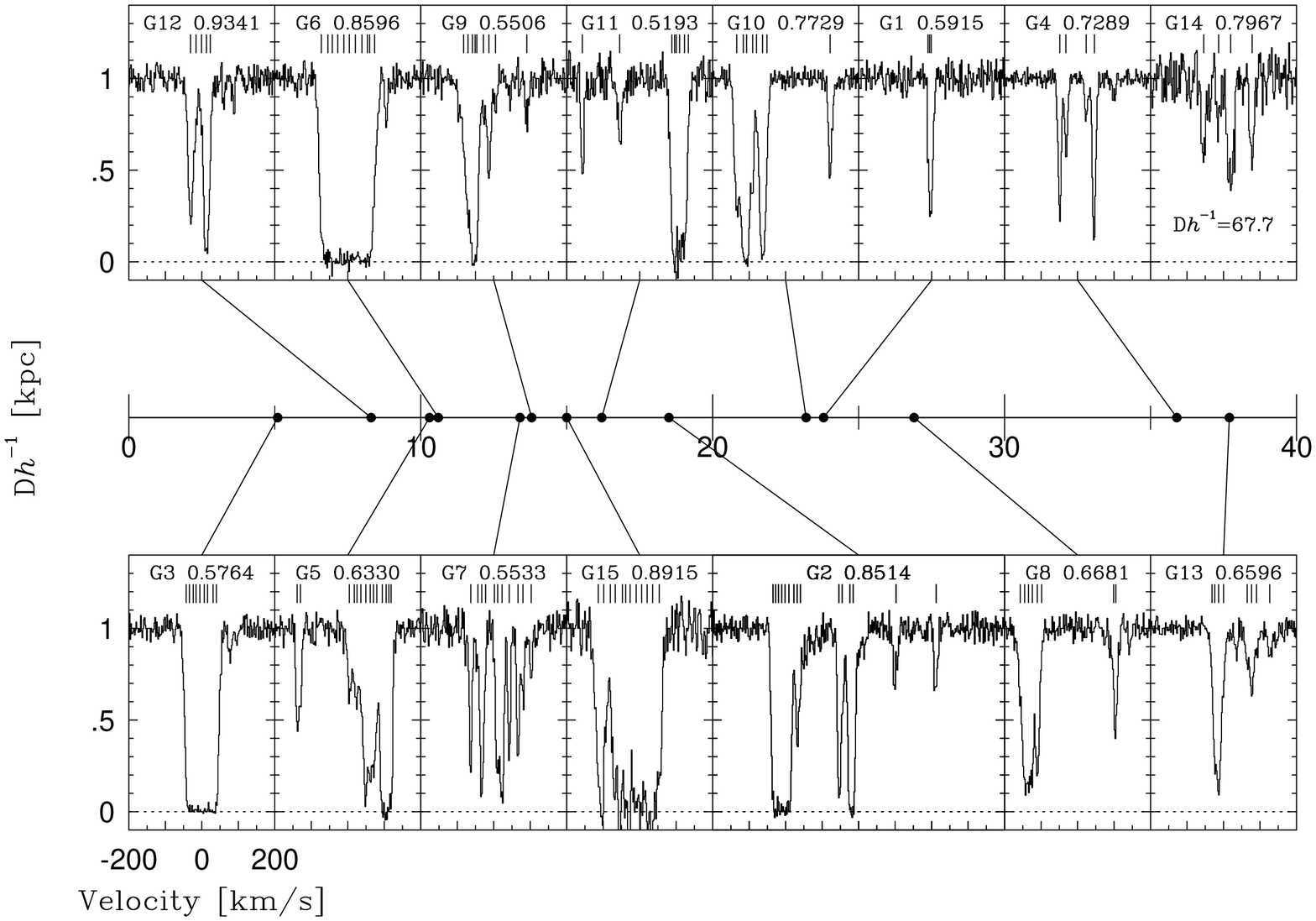}                                                
\caption
{The {\Mg} $(\lambda 2796)$ absorption profiles obtained with HIRES on
the Keck telescope plotted in order of increasing impact parameter, 
$Dh^{-1}$~kpc, of the associated absorbing galaxy.  
The vertical ticks above the continuum of each absorption feature
mark the subcomponents used in the analysis.}
\end{figure}                                                    
                                                                
The median is used in lieu of the average because the former is
symmetric about the number of clouds and is not sensitive to 
one or a few large $\Delta v_{i}$ clouds.
The second kinematic indicator is the number of absolute
deviations of the highest velocity cloud,
\begin{equation}
N_A = \frac {{\rm max}(\Delta v^{+},|\Delta v^{-}|)}
            {A({\Delta v})} ,
\end{equation}
where $\Delta v^{+}$ and $\Delta v^{-}$ are the largest redshifted and
blueshifted cloud velocity differences from the system median,
respectively.
This quantity, which ranges between $1 \leq N_A \leq 4$ for our data,
measures the degree and presence of an extreme kinematic ``outlier''
in each system.
The third kinematic indicator is a measure of the kinematic asymmetry,
\begin{equation}
\eta  = \frac {\min(\Delta v^{+},|\Delta v^{-}|)}
              {\max(\Delta v^{+},|\Delta v^{-}|)} ,
\end{equation}
where $\eta = 1$ indicates that the extreme redshifted and blueshifted
clouds lie symmetrically about the system median, and small $\eta$
indicates high asymmetry.

\section{Testing For Correlations}

We ran non--parametric Spearman and Kendall correlation tests on
the measured galaxy and absorption properties, which are presented in
Table 2.
The Kendall tests better handle the treatment of data in which a
value appears multiple times (fractional ranks), as is the case with
$N_{\rm c}$.
Galaxies for which a given property was not measured were excluded
from the test pertaining to the missing property.
Using the criterion $P_{\rm k} \leq 0.01$ (99\% probability that
no correlation is not consistent with the data) to indicate the
presence of a correlation, we found {\it no}\/ statistically
significant correlations between the absorbing gas strengths or
kinematics with the galaxy properties.
To examine the effects due to projected galactocentric distance, $D$,
and due to path length through the halo, $S = 2\sqrt{R^{2}-D^{2}}$,
where  $R = 38h^{-1}(L_K/L_K^{\ast})^{0.15}$ (\cite{s95}), we ran
additional tests on the absorption properties versus $D/R$ and $S$,
and versus $L_B$ and $L_K$ normalized by $D$ and by $S$. 
One might expect correlations to be revealed from the normalized $L_K$
if absorbing gas exhibited a smooth radial spatial distribution
that scales with the luminosity to some power, as expected for
pressure--confined clouds in halos where the pressure gradient scales
with galaxy mass (\cite{mo96}).
Again, we found no correlations.

To investigate sensitivity to small--number statistics, and to
ascertain if our sample of 15 galaxies is a fair representation of the
SDP sample, we also performed Monte--Carlo rank correlation tests. 
For each simulation, we assigned the observed absorption properties
(columns 8--13 in Table 2) to 15 galaxies drawn at random from 54 $z \leq
1 $ SDP galaxies. 
For each correlation test, 5000 of these simulations were executed and
the probability $P_{\rm MC}$ that the measured $\tau_{k}$ or greater
would be measured from randomized galaxy--absorption properties for
our sample was computed.
The $P_{\rm MC}$ strongly correlate with the $P_{\rm k}$, which is
indicative that our sample is a representative subset of the SDP data set
and that the measured $P_{\rm k}$ are reliable indicators.

In Table 3, we present those tests for which both the simulated and
measured $P$ are weakly suggestive of correlations in that the null
hypothesis (no correlation) is not consistent with the data at the
2$\sigma$ confidence level ($P_{\rm k} \leq 0.1$).
Since one or a few kinematically simple or complex systems could be
dominating a test for which a correlation is suggested at the
$\sim$~2$\sigma$ level, we re--ran the Spearman and Kendall tests on
our sample, each time omitting a single galaxy.   
We then examined the {\it distribution}\/ of $P_{\rm k}$ for each test
to see if a large $P_{\rm k}$ results from the omission of any one
galaxy. 
For the correlation tests tabulated in Table 3, the $P_{\rm k}$
distributions cluster tightly about the their respective values,
except for $A(\Delta v)$ versus $S$ (note the discrepancy between
$P_{\rm k}$ and $P_{\rm MC}$ for this case).

\section{A Second Look at the Galaxy/Halo Model}

We further investigated the ``halo'' model described by Lanzetta \& Bowen
(1990, 1992).
As stated above, the primary test of this model is the prediction that
the observed differences in the absorption properties from one
system to another are predominantly due to the QSO--galaxy impact
parameter.
Thus, we performed maximum likelihood least squares linear fits (LSFs)
to the quantities tested against impact parameter that are listed in
Table 3.
The goal is to obtain the significance level of each fitted slope.
If any of these LSF slopes is not inconsistent with zero, within the
uncertainties applied to the data, then we may have further evidence
supportive of a possible correlation.

\begin{figure}[bh]                                               
\plotone{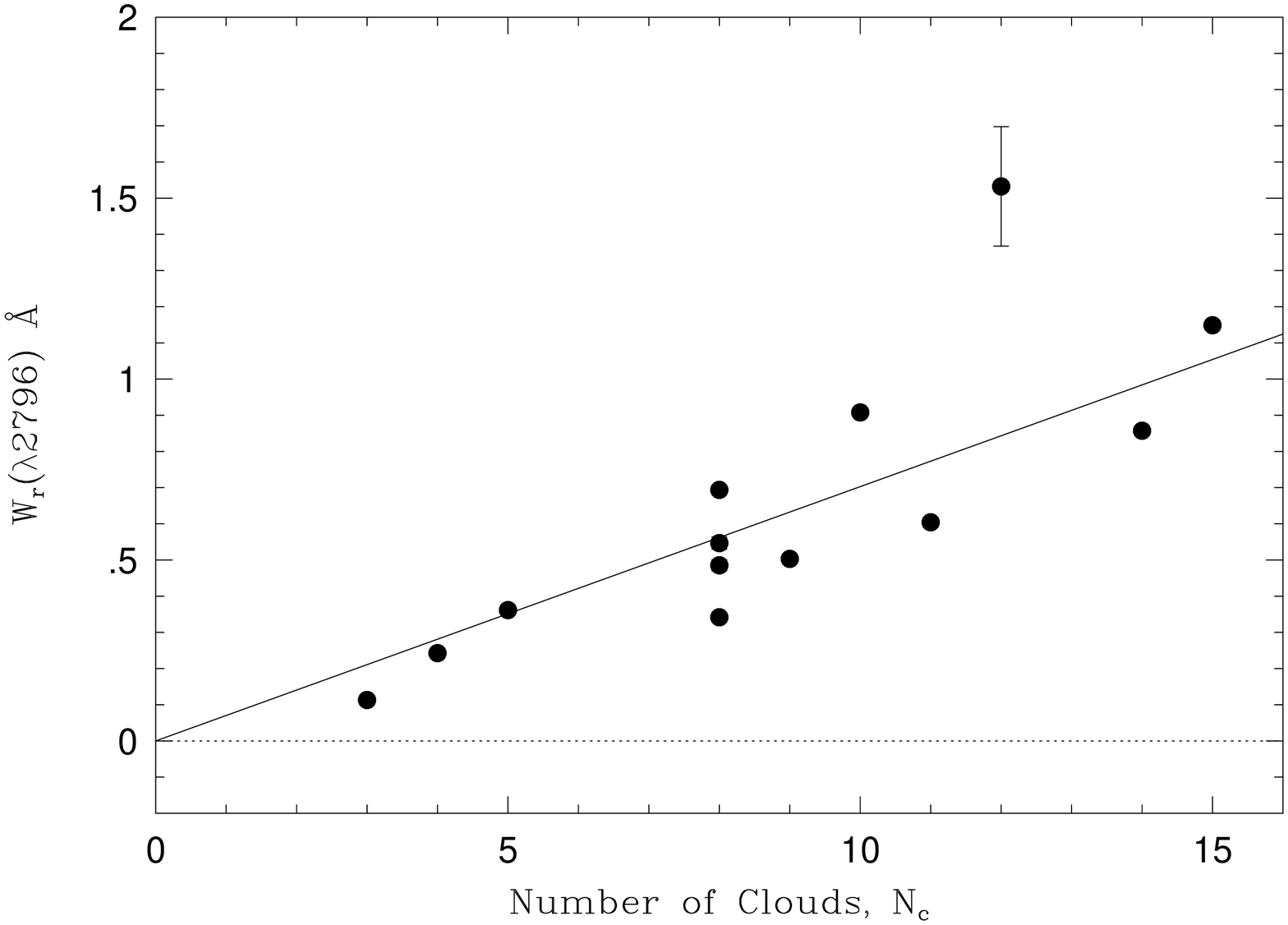}                                                  
\caption                                                          
{The rest equivalent width of {\Mg} $(\lambda 2796)$ versus
the number of clouds for 13 of the systems presented in Table 2.  
The error bars are the formal errors from the
spectral analysis and are roughly the size of the data points, with
the exception of that measured for G15. The LSF slope has been forced
through the origin.}
\end{figure}                                                      

For the LSF fits, we modeled the uncertainty of each absorption
property as scatter due to Poisson fluctuations, which are predicted
from the fact that the QSO line of sight samples a finite number of
clouds in each galactic system. 
For the equivalent widths [cf.~equation 2 of Lanzetta \& Bowen
(1990)], the LSF uncertainty is given by,
\begin{equation}
\sigma = \left\{ \sigma ^2(W_{\rm r}) + k^{2}[W_{\rm r}/k +1]
\right\} ^{1/2},
\end{equation}
where $\sigma (W_{\rm r})$ is the measured uncertainty in $W_{\rm
r}$ and the ``scatter'' term follows from Poisson counting statistics
if the $W_{\rm r}$ are correlated with $N_{\rm c}$, following $W_{\rm
r} = kN_{\rm c}$ (\cite{pb90}).
We obtained the LSF relation $k=0.0703\pm0.0002$ to the data presented
in Figure 3.
The value of $k$ may actually depend upon redshift, but can be 
approximated as a constant over $0.5 \leq z \leq 0.9$, since $W_{\rm
r}$ and $N_{\rm c}$ for our sample do not correlate with $z$. 
If unresolved sub--components remain hidden in the HIRES profiles, our
measured $k$ would be an upper limit.  
A smaller $k$ would result in smaller modeled scatter.

In Figure 4, we plot the total absorption strength, $W_{\rm r}$, verse
impact parameter, D$h^{-1}$~kpc.
The solid circles are taken from Table 2, where the error bars are
the modeled scatter as described above.
The open circles are the full $z\leq1$ SDP data set, which we
presented to illustrate that our subsample appears to be a fair
representation of the full available sample.
Applying the modeled scatter as uncertainties in $W_{\rm r}$, we found
that an LSF to the relation $W_{\rm r} = aD^{\alpha}$ yields $\alpha =
-0.28\pm0.50$ with $\chi^{2}_{\nu} = 1.3$.
Thus, we found no anti--correlation at the 99.9\% confidence level of
$W_{\rm r}$ with $D$ for our 15 systems.
Apparently, the spatial number density of ``halo'' absorbing clouds
does not follow a simple power law with galactocentric radius.
For comparison, we also show the LSF fit obtained by Lanzetta \& Bowen
(1990) to a similar size but different data set (not shown).
Our LSF fits to DR~$= \alpha D + A$ and $N_{\rm c} = AD^{\alpha}$ both
yielded slopes statistically consistent with $\alpha = 0$, where we
have assigned statistical scatter (uncertainties) to DR and $N_{\rm
c}$ consistent with those of the $W_{\rm r}$.

\begin{figure}[bh]                                                
\plotone{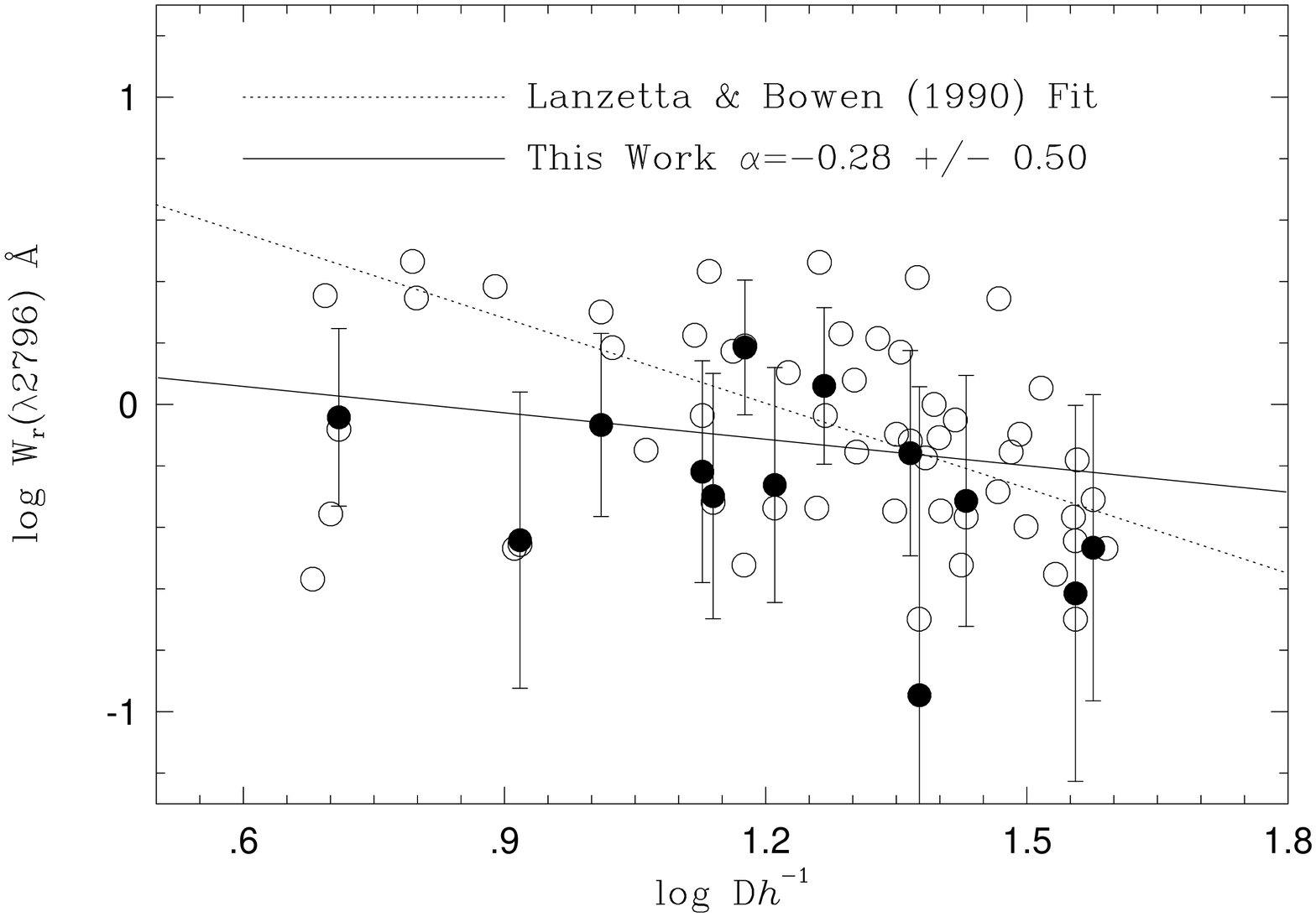}                                                  
\caption                                                          
{$W_{\rm r}$ $(\lambda 2796)$ versus the QSO--galaxy impact parameter
$D$. The open circles are the $z\leq 1$ SDP dataset and the filled
circles are the 13 systems measured in this work.
The error bars are the modeled scatter in the $W_{\rm r}$ due to
intercepting a finite number of clouds along the line of sight.
The dashed line shows the LSF slope to the assumed relationship $W_{r} =
AD^{\alpha}$ found by Lanzetta \& Bowen (1990) using a smaller and
different data set of {\Mg} absorbers.  The solid line is our LSF result.
We found $\alpha = -0.28\pm 0.50$, which is not a statistically
significant slope.}
\end{figure}                                                      

It is important to consider whether non--unique solutions for the
profile fits, as illustrated in Fig.~1, could significantly effect our
conclusions.
Consider the number of clouds, $N_{\rm c}$, which is a very central 
quantity in that it either directly or indirectly effects the computed
values of the kinematic indicators.
Our fitting philosophy was to introduce the {\it minimum}\/ number of
subcomponents that yielded model spectra consistent with the
constraints provided by the error spectra.
Thus, the $N_{\rm c}$ are lower limits, since finite signal--to--noise
and instrumental resolution could conceal narrow subcomponents.
The modeled Poisson fluctuations in the $N_{\rm c}$ are typically
a factor of a few larger than the uncertainty from the profile fits,
though this is difficult to precisely quantify.
Our results would be altered only if there was some unknown systematic
error in the fitted $N_{\rm c}$ with the impact parameter, $D$, {\it
and}\/ if the uncertainty in the fitted $N_{\rm c}$ were roughly the
same size as the modeled scatter.  

That we have found no dependence of $N_{\rm c}$ with $D$ has
implications for predictions from systematic rotational and radial
flow kinematic models.  
In particular, consider the median absolute deviation, [$A(\Delta v)$,
equation 1] versus QSO--galaxy impact parameter, which we have plotted
in Figure 5.
Since the number of clouds is independent of $D$, we can approximate
$A(\Delta v) \propto \left< \Delta v \right> $, the mean $\Delta v_i$.
For infall models, $\left< \Delta v \right> \propto (1-D^2/R^2)^{1/2}$,
and for rotational models, $\left< \Delta v \right> \propto 1 - D/R$,
where $R$ is the ``halo'' or ``disk'' size.
Both models predict that $A(\Delta v)$ decreases with $D$ and sharply
cuts off at $D=R$.
There is no such anti--correlation in our sample.
In fact, the Spearman and Kendall tests are weakly suggestive of a
correlation, not the predicted anti--correlation.
In particular, it is difficult to reconcile either of these kinematic
models with the $D\sim 70$~kpc absorption observed for galaxy G14
(cf.~\cite{cc96}), though such an occurrence is likely to be fairly
rare (based upon $dN/dz$ constraints, \cite{steidel93b}).
For the case of rotation in disk galaxies, it is likely that for
$D\leq 20$~kpc, the random inclination of disk galaxies would
introduce a sizable scatter in the $A(\Delta v)$.
We note, however, that this scatter would always be toward smaller
$A(\Delta v)$ values as disk inclination decreases, and thus there
would still be an $A(\Delta v)$ versus $D$ upper envelope that does
exhibit an anti--correlation.

\begin{figure}[bh]                                                
\plotone{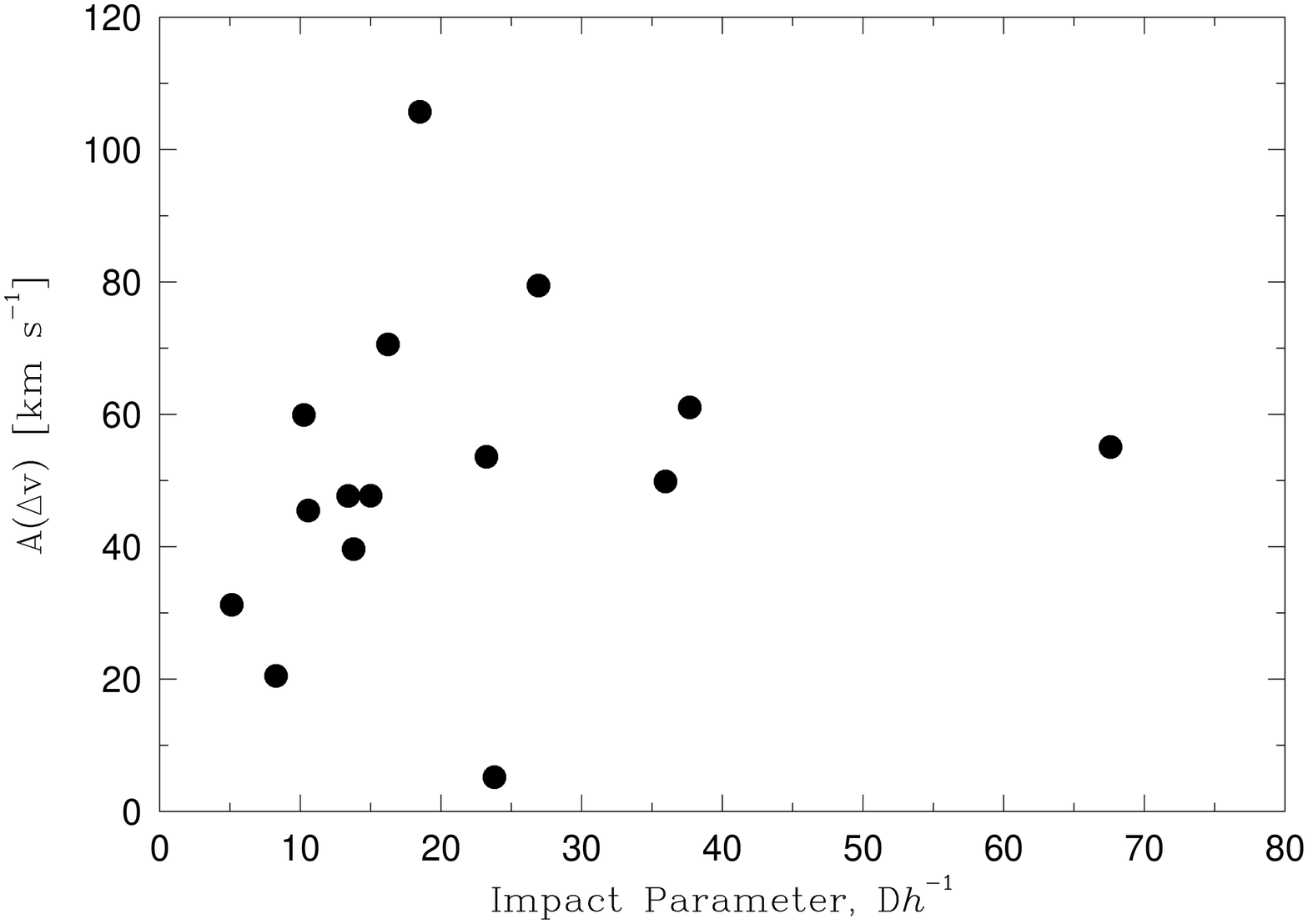}                                                  
\caption                                                         
{The median absolute deviation of cloud velocities (equation 1) versus
the QSO--galaxy impact parameter.  There is no clear trend in the 
line--of--sight velocity ``dispersion'' with the projected galactocentric
distance probed by the QSO light path.  Both the rotational and infall
kinematic models of galactic halos predict a decreasing spread with
impact parameter (see text).}
\end{figure}                                                      
                                                                 
Overall, what is clear, is that the absorption properties of {\Mg}
absorption--selected galaxies exhibit a level of scatter greater than
that predicted by simply assuming Poisson fluctuations on the grounds
that each line of sight is sampling a finite number of clouds.
Based upon 15 systems, we cannot conclude (with confidence) that weak
correlations do or do not exist between the quantified 
kinematic properties of the absorbing gas and the galaxy properties
available from ground--based images, but we can argue, based upon the
above discussion, that ``smooth'' correlations can be ruled out.
{\it Of primary significance is the fact that the QSO--galaxy impact
parameter apparently does not provide the primary distinguishing
factor by which absorption properties can be characterized}.

The implications are that (1) the spatial distribution of absorbing gas
surrounding intermediate redshift galaxies is not smoothly varying,
and that (2) the velocities of the absorbing gas clouds cannot be
described by a {\it single} systematic kinematic model.  
One may infer that these quantities may not be roughly identical
from galaxy to galaxy of similar morphological type, even if the
events and processes that give rise to the gas are.
Such a picture is difficult to reconcile with theoretical results
that predict a ``steady state'' in which cloud coalescence and growth
from the cooling of hot gas in a virialized halo is balanced by cloud
disruption and evaporation, giving rise to a near unity covering factor
(cf.~\cite{mo96}; \cite{anninos95}; \cite{begelmann90}).
A ``steady state'' halo of this type would likely show systematic trends
with less scatter than is observed, and is not well supported by
observations of local galaxies (\cite{bbp95}), provided there is no
rapid halo evolution from $z \sim 0.4$ to the present.

\section{Discussion}

If {\Mg} absorption profiles contain spectroscopic imprints of
systematic spatial and velocity distributions of the low ionization
gas comprising disks and halos, as suggested by Lanzetta \& Bowen
(1992), it appears from our results that absorbing gas would have to
arise from a heterogeneous population of events and mechanisms that
each impart their own systematic kinematic signatures, and possibly
chemical and ionization conditions.
{}From lines of sight through the Milky Way, the Large Magellanic
Cloud (LMC), and local galaxies, a substantial variation in the
absorption properties corresponding to a variety of dynamical events
has been observed (\cite{sav93}; \cite{lu94}; \cite{car95};
\cite{sem95}; \cite{bbp95}; \cite{caulet96}).
In the vicinity of a ``typical'' {\Mg} absorption--selected galaxy,
such events would need to be on--going over a Hubble time in order to
explain the sustenance of extended gaseous regions [as inferred from
their non--evolving co--moving cross section (\cite{ss92})], since the
absorbing cloud lifetimes are $\sim 1$~Gyr (\cite{lb90}; \cite{mo96}).

Hubble Space Telescope (HST) WFPC2 images of $z \sim 1$ galaxies from
the DEEP Project reveal a diversity of morphological types, including
small compact galaxies, and apparently normal late--type spirals
(\cite{koo96}).
These images also reveal evidence suggestive of a higher frequency of
merging in the past and for interactions of small, gas--rich galaxies with
each other or with larger, well--formed galaxies.
Apparently, a mixture of physical processes is at work in the formation
and evolution of field galaxies.
Based upon local studies, Bowen, Blades, and Pettini (1995) have
suggested that recent merging events may play an important role in the
presence of {\Mg} absorbing gas, and caution against a generalized
halo model.

On the other hand, one cannot completely rule out the idea that some
fraction of the observed systems are dwarfs (i.e.~similar to Carina,
Leo I, etc.) that undergo episodic self--regulating star formation,
resulting in the blow out of gas to large radii over a Hubble
time. 
Occasionally, a gas rich low mass galaxy could lie within $D\sim
50$~kpc and $\Delta v \sim 100$~{\kms} of an ``identified'' galaxy by
virtue of projection, especially in the instances when ``double
galaxies'', which are suggestive of Local Group environments, are seen
to be associated with {\Mg} and/or {\C} absorption.
However, it is difficult to reconcile a large contribution from this
latter picture due to an observed correlation of the absorption
cross--section with the identified galaxy $K$ luminosity (\cite{s95}).
This correlation suggests that galactic mass has some connection
with a galaxy's ability to organize tidally stripped, accreting, or
infalling material (cf.~\cite{mo96}; \cite{s95}).
The important point here is that unbiased deep imaging and thorough
spectroscopic surveys of QSO fields are still in progress, and as
shown by Charlton \& Churchill (1996b), the simple picture of a
near--unity covering factor ``galactic halo'' is not yet conclusively
demonstrated.

Even though models of the spatial distributions and kinematics of the
{\Mg} absorbing gas around galaxies are still not well constrained by
the observations, there is little doubt that the galaxies identified
with {\Mg} absorption are indeed associated in some way with the
processes that give rise to the absorbing gas (SDP; \cite{s95}).
As such, the types and frequencies of the {\it events}\/ that give
rise to absorbing gas are likely to show trends with galaxy
morphologies and impact parameter, even if there is a great deal of
scatter in these trends.
It remains to be understood to what degree extended and warped {\HI}
disks in spiral galaxies contribute to absorption at all impact
parameters ({\cite{cc96}), though it is highly probable that they
actually dominate for $D \leq 20h^{-1}$~kpc and perhaps can even
account for most strong saturated components out to $D \leq
40h^{-1}$~kpc.
It is interesting to note the extended and complex structure of high
column density {\HI} gas surrounding M81, which is seen to be
distributed in a flattened geometry and is associated with {\it
current}\/ satellite mergings (\cite{yun94}).
This observation also supports the idea that for $r_{\rm gal} \geq
20$~kpc and for the range of observed morphological types, likely
sources of gas are the tidal stripping of or blow out from orbiting
dwarf satellite galaxies (\cite{wang93}), the presence of dwarf
galaxies very close to the QSO line of sight (\cite{york86}), or the
accretion of clumps (fragments) of intergalactic gas (\cite{mo96}).
Since gas producing events (such as superbubbles) from both galaxies
and dwarf companions give rise to similar absorption strengths and
kinematics over a $\sim 50$~kpc galactocentric range of impact
parameters [as seen for the Milky--Way and environs (\cite{lu94};
\cite{caulet96}; \cite{welty96})], it will be difficult to interpret
the origins of absorbing gas until deep high--spatial resolution images
of the galaxies can be incorporated into QSO absorption line studies.

Based upon the available data, one may tentatively infer that {\Mg}
absorption--selected galaxies are roughly identical in the sense that
they are continually processing and accreting gas, but differ in that
the spatial and kinematic distribution of the gas {\it around}\/ each
galaxy depends upon the events recent to the epoch at which it is
observed.
It is likely that on--going gas processing events are continuously
reorganizing sub--galaxy distributions of gas on timescales of a few
Gyr.  
This is {\it not}\/ to imply that global reorganization of gaseous
kinematic, ionization, and spatial distributions occurs in all {\Mg}
absorbing galaxies.
We caution that the data do {\it not}\/ rule out the idea that
sub--galaxy distributions are somewhat stable but have a high level of
structure, since it is possible that observations of a single galaxy
and its surroundings along different lines of sight at a single epoch
in its evolution could give rise to the wide variety of observed
absorption profiles.
A test of these issues would include establishing which, if any, of
the following observed variations in absorption properties show some
type of systematic trend: (1) galaxies of different morphological
types, (2) similar lines of sight through galaxies of a given
morphological type at different cosmic epochs, or (3) different lines
of sight through galaxies of a given morphological type at a given
cosmic epoch.  

For example, statistical differences may exist between samples of
profiles for the different morphological types.  
This might imply that the relative importance of the various
on--going processes that produce and adjust the distribution of gas is
different in the different morphological types.
Similarly, there could be statistical differences between samples of
profiles for galaxies of the same morphological types at different
redshifts.  
This might be indicative that the relative importance of the various
on--going processes is changing.
Since ellipticals and spirals have distinctive internal stellar
dynamics, their gas kinematics may also be distinctive, especially for
$r_{\rm gal} \la 20$~kpc.   
For both ellipticals and spirals, if the events at larger
galactocentric radius arise predominately in satellite galaxies and/or
are stochastic accretion events lasting $\sim$ few Gyr, they will be
observable for the full range of observable $D$.
Thus, the line--of--sight quantities $N_A$ and $\eta$ would be expected
to exhibit scatter with $D$, since they may be sensitive to the chance
interception of ``active'' halo material.
As such, inferring the distinction between kinematic systematics based
upon morphological type at small $D$ would be complicated. 
It will be an interesting exercise to try to separate out the
absorption properties by galaxy morphology once the number of observed
{\Mg} absorbing galaxies becomes large. 

We are obtaining deep WFPC2 images of the QSO fields with
the goal of obtaining further clues for interpreting the kinematics
measured from HIRES profiles.
In the case of spiral galaxies, inclination and orientation with
respect to the line of sight may play an important role.
It is interesting to compare G7, a face--on spiral, and G15, a highly
inclined spiral, which have comparable $L_B$ luminosities, 
$N_{\rm c}$, $N_A$, and $D$, but slightly different $K$ luminosities,
and strikingly different absorption profile shapes.
G7 exhibits lower column density clouds that are kinematically
symmetric about the galaxy $z$ direction, whereas G15 exhibits 
higher column densities.
The galaxy G2, which is a compact blue diskless galaxy, exhibits
what appears to be a great deal of recent kinematic activity (see
Fig.~1) and a remarkably more complex absorption than is seen in the
spirals G7 and G15.
Perhaps the line of sight through G2 is sampling a recent and large
accretion event.

\section{Conclusions}

As a first step toward developing a better appreciation of the
spatial distributions and kinematics of low ionization absorbing gas
associated with galaxies, we have presented the {\Mg} $(\lambda 2796)$
absorption profiles for 15 $z < 1$ galaxies, developed simple
kinematic indicators of the profiles, and searched for statistically
significant correlations between the absorbing gas and galaxy
properties.

\begin{enumerate}
\item
Using Spearman and Kendall non--parametric rank correlation tests,
we found no evidence for 2.5$\sigma$--level correlations between
the absorbing gas and galaxy properties.
We performed Monte--Carlo Spearman and Kendall tests in which we
randomly assigned our measured absorption properties to the full
SDP sample of galaxies.
These tests revealed that our 15 galaxies are a fair
representation of the SDP dataset and that the Spearman and Kendall
results provide reliable indicators for the presence or non--presence
of correlations.
\item
In Table 3, we have listed the eight tests for which the Spearman and
Kendall results are weakly suggestive of correlations. 
Since trends in absorption properties with the QSO--galaxy impact
parameter, $D$, are an important test of the galaxy/halo model, we
have performed maximum likelihood least square fits to those tests
that included $D$ (i.e.~DR, $N_{\rm c}$, and $W_{\rm r}$ versus $D$).
For these fits, we modeled the uncertainty in the measured absorption
properties as scatter due to Poisson fluctuations, which are predicted
when sampling a finite number of absorbing clouds along the line of
sight through a galaxy halo.
All LSF slopes were consistent with zero, which is indicative of no
statistically significant dependence with $D$ for these absorption
properties.
\item
Since the number of clouds shows no significant dependence with $D$,
both the rotation and radial flow kinematic models predict an
anti--correlation of the median absolute deviation of cloud velocities
with impact parameter.
As illustrated in Fig.~5., we found no clear trend in $A(\Delta v)$
with $D$.
If anything, the weak positive correlation suggested by the rank
correlation tests is suggestive of a scenario in which more than one
kinematic model may be needed to explain the data.
Perhaps the inner 20~kpc are dominated by the systematic kinematics 
expected for galaxies (morphology dependent?), whereas the higher
impact parameter kinematics is dominated by accretion and merging of
various types of ``halo'' material.
\item
We have found that the QSO--galaxy impact parameter apparently does
not provide the primary distinguishing factor by which absorption
properties can be characterized. 
This fact suggests that we should investigate the degree to which
galaxy morphologies and, in the case of disk galaxies, orientations
with respect to the QSO light path play a role in distinguishing
between observed absorption properties.
\item
Since {\Mg} absorption properties exhibit a level of scatter greater
than that predicted by a simple Poisson fluctuation model, our data do
not provide a level of constraint necessary for us to infer
functional relations that describe the spatial and kinematic
distributions of {\Mg} absorbing gas around galaxies.
If weak correlations are in fact present, they exhibit a large enough
scatter that a fair sample of 15 systems has not yielded unequivocal
results.
\item
As one possible interpretation, we have tentatively suggested a
picture of intermediate redshift galactic halos in which the
distribution and kinematics of the low ionization absorbing gas is
dominated by the events and mechanisms that give rise to the gas
recent to the epoch of observation.
These halos would be dynamically active with evolving gaseous conditions
dependent upon local environmental influences, as expected with
tidally stripped dwarf satellite galaxies, infalling gaseous
sub--galactic fragments, superbubbles, and in the case of disk
galaxies, with high velocity clouds, galactic fountains, and the
warped extended disks themselves.
Our observational results are not suggestive of a picture in which
galaxy halos underwent a single epoch of dynamical formation in the
past.
Rather, these galaxies have experienced multiple and episodic merging
and have undergone internal events that give rise to extended (though
somewhat patchy) regions of low ionization absorbing gas.
\item
If the {\Mg} absorbing gas from these events has a dynamical time of
$\sim$ 1 Gyr, the gas producing events, spread out both spatially and
temporally, would not give rise to halos comprised of multiple
coalescing and evaporating absorbing clouds that yield a near--unity
covering factor.
Such halos would be expected to exhibit systematic kinematics and have
an approximately spherically symmetric spatial number density
distribution that scales with galactocentric radius.
Instead, these events may give rise to $\sim$~few Gyr sub--galactic
gaseous structures so that the overall spatial distribution and
kinematics of absorbing gas in any one galaxy would be expected to
change over a few cloud dynamical times.
The important point is that the data are not suggestive of the continual
processing of gas that would give rise to an apparently ``steady
state'' halo over the 10~Gyr evolution of the galaxy.
\item
Since the dynamical time is roughly an order of magnitude shorter than
the overall look--back time over which complex {\Mg} absorption is seen
around galaxies, and since QSO absorption lines likely sample gas
produced in events recent to the epoch at which it is observed, we
should be able to directly track evolution in the absorption properties.
A larger database of HIRES {\Mg} profiles would be instrumental for
directly quantifying evolution in the absorbing gas properties over
the redshift range $0.4 \leq z \leq 2.2$.
Such evolution may provide information helpful for ruling out 
various type of events or absorbing structures and for quantifying
possible major shifts in epochs from one type of pre--dominant 
galaxy/halo gas processing phase to another.
\end{enumerate}

A more comprehensive appreciation of the spatial and kinematic
properties of low ionization absorbing will require a sizable unbiased
sample of high resolution absorption profiles (unbiased in that the
distribution of galaxy luminosities, colors, impact parameters, and
absorption rest equivalent widths are consistent with having been
drawn from the observed distributions of {\Mg} selected galaxies).
It is important that surveys of QSO fields be complete to a small
limiting rest frame luminosity and that the selection of these fields 
be unbiased with regard to the presence or non--presence of
intervening absorbing galaxies.

Unless the gas kinematics in spiral/disk and elliptical galaxies are
distinguishable as seen in high resolution profiles, it is not likely
the {\Mg} absorption properties can be used to infer the type of
galaxy associated with the absorbing gas at the highest redshifts,
where the galaxy properties are difficult to obtain. 
However, we may be able to roughly infer the region of the galaxy
sampled by the line of sight or something about its current or recent
environmental dynamical activity.
Ultimately, the observed evolutionary properties of gas derived from
QSO absorption line studies are likely to yield direct quantities from
which the  formation and evolution time scales of
10$^{12-13}$~M$_{\sun}$ structures in the universe can be better
understood.

\acknowledgments
This work was supported in part by the California Space
Institute grant CS--1194, NASA grant NAGW--3571, and NSF grant
AST--9457446.
CCS thanks the Alfred P.~Sloan Foundation.
CWC and SSV expresses sincere appreciation to Mike Keane for his
assistance in getting this research program started and for his
assistance with one of the observing runs.
Further thanks go to Wayne Wack and Joe Killian for their efficient 
operation of and assistance with the Keck I telescope.
It is a pleasure to acknowledge A.~Boksenberg, M. Dickinson D. Meyer,
P. Petitjean, and D. York for stimulating discussions, and especially
J. Charlton, R. Guhathakurta, G. Smith, and D. Zaritsky for feedback
on ideas that appear in this work.  Additional thanks to J. Charlton
for her critical reading of an earlier form of this manuscript.

\begin{deluxetable}{lcccrcc}
\tablewidth{0pc}
\tablewidth{320pt}
\tablecaption{\sc Journal of HIRES/Keck Observations}
\tablehead
{
\colhead{QSO} & 
\colhead{~V [mag]} & 
\colhead{~$z_{\rm em}$} &  
\colhead{~Exp [s]} &
\multicolumn{2}{c}{$W_{\rm min}(\lambda 2796)$ [\AA]} \\
\cline{5-6} 
}
\startdata
$0002+051$  & 16.2 & 1.899 &  2700 & 0.005 & 0.004 \nl 
$0117+212$  & 16.1 & 1.491 &  5400 & 0.011 & 0.009 \nl
$0420-014$  & 17.0 & 0.915 &  3600 & 0.016 &       \nl
$0454+036$  & 16.5 & 1.343 &  4500 & 0.007 &       \nl  
$1148+384$  & 17.0 & 1.299 &  5400 & 0.018 &       \nl 
$1222+228$  & 15.5 & 2.040 &  3600 & 0.016 &       \nl 
$1241+174$  & 15.4 & 1.282 &  2400 & 0.008 &       \nl 
$1248+401$  & 16.3 & 1.032 &  4200 & 0.005 &       \nl  
$1254+044$  & 16.0 & 1.018 &  2400 & 0.017 & 0.009 \nl   
$1317+274$  & 16.0 & 1.014 &  3600 & 0.009 &       \nl 
$1622+235$  & 18.3 & 0.927 & 19240 & 0.023 & 0.020 \nl
\tablecomments
{The quoted exposure time is the sum of combined observations.
The quantity $W_{\rm min}$ is the 5$\sigma$ detection rest
equivalent width limit of a feature in the velocity range $\pm
600$~{\kms} from the absorber redshift.  For the cases where two
absorbers lie along the QSO line of sight, the second entry
corresponds to the higher redshift system.
}
\enddata
\end{deluxetable}

\begin{deluxetable}{ccccccccccccc}
\tablewidth{0pc}
\tablecaption{\sc Absorbing Galaxy Properties}
\tablehead
{
& & \multicolumn{5}{c}{Galaxy Properties} &
\multicolumn{4}{c}{Absorption Properties} \\
\cline{3-7} \cline{8-13} 
\colhead{ID} &
\colhead{QSO} & 
\colhead{$z_{\rm gal}$} &  
\colhead{$L_B$} & 
\colhead{$L_K$} & 
\colhead{$B-K$} &
\colhead{$Dh^{-1}$} & 
\colhead{$W_{\rm r}$} &
\colhead{DR} &
\colhead{$N_{\rm c}$} &
\colhead{$A({\Delta v})$}  &
\colhead{$N_{A}$} &
\colhead{$\eta$}  \\
\colhead{(1)} &
\colhead{(2)} &
\colhead{(3)} &
\colhead{(4)} &
\colhead{(5)} &
\colhead{(6)} &
\colhead{(7)} &
\colhead{(8)} &
\colhead{(9)} &
\colhead{(10)} &
\colhead{(11)} &
\colhead{(12)} &
\colhead{(13)}
}
\startdata
 G1 & 0002 & 0.5914 & 1.09 & 1.34 & 4.12 & 23.8 & 0.11 & 1.51 &  3 &   5.2 & 1.06 & 0.89 \nl
 G2 &      & 0.8514 & 1.23 & 0.47 & 2.86 & 18.5 & 1.14 & 1.11 & 15 & 105.7 & 3.60 & 0.17 \nl 
 G3 & 0117 & 0.5763 & 2.32 & 2.54 & 4.00 &  5.1 & 0.91 & 1.07 &  9 &  31.2 & 2.10 & 0.77 \nl 
 G4 &      & 0.7289 & 3.27 & 3.65 & 4.02 & 36.0 & 0.24 & 1.67 &  4 &  49.9 & 1.00 & 0.90 \nl
 G5 & 0420 & 0.6330 & 0.28 & 0.34 & 4.11 & 10.3 & 0.86 & 1.32 & 14 &  59.9 & 3.24 & 0.32 \nl  
 G6 & 0454 & 0.8596 & 0.26 &      &      & 10.6 &      &      & 11 & 45.5 & 1.68 & 0.91 \nl   
 G7 & 1148 & 0.5532 & 0.75 & 0.56 & 3.59 & 13.4 & 0.60 & 1.61 & 11 &  47.7 & 1.95 & 0.79 \nl   
 G8 & 1222 & 0.6681 & 0.49 & 0.45 & 3.80 & 26.9 & 0.48 & 1.51 &  8 &  79.5 & 2.80 & 0.18 \nl   
 G9 & 1241 & 0.5507 & 0.64 & 0.41 & 3.42 & 13.8 & 0.50 & 1.37 &  9 &  39.6 & 3.48 & 0.25 \nl   
G10 & 1248 & 0.7725 & 0.53 & 0.27 & 3.15 & 23.2 & 0.69 & 1.25 &  8 &  53.6 & 3.87 & 0.23 \nl   
G11 & 1254 & 0.5192 & 0.14 & 0.11 & 3.65 & 16.2 & 0.55 & 1.49 &  8 &  70.6 & 3.61 & 0.14 \nl   
G12 &      & 0.9341 & 0.43 & 0.26 & 3.35 &  8.3 & 0.36 & 1.51 &  5 &  20.5 & 1.56 & 0.77 \nl   
G13 & 1317 & 0.6598 & 2.30 & 2.10 & 3.80 & 37.7 & 0.34 & 1.55 &  8 &  61.0 & 1.55 & 0.66 \nl  
G14 & 1622 & 0.7967 & 1.63 & 0.74 & 3.04 & 67.6 &      &      &  4 &  55.1 & 1.36 & 0.77 \nl
G15 &      & 0.8915 & 0.71 & 0.33 & 3.07 & 15.0 & 1.53 & 1.12 & 12 &  47.7 & 1.79 & 0.95 \nl
\tablecomments
{Column 1 provides the galaxy ID number.
Listed in column 2 are the first four digits which designate the QSOs
listed in Table 1.  The galaxy's redshift, and $B$ and $K$ luminosities
(in terms of $L_{B}^{\ast}$ and $L_{K}^{\ast}$, respectively) are
listed in columns 3, 4, and 5.  The rest--frame $\left< B-K\right>$
colors are listed in column 6, and the impact parameters in kpc are
given in column 7.   The absorption properties, measured from the
HIRES spectra, include the rest--frame $\lambda 2796$ equivalent width
(column 8) and the doublet ratio DR~$=W(\lambda 2796)/W(\lambda 2803)$
(column 9).  Columns 10 and 11 list the number of clouds $N_{\rm c}$
and the mean absolute deviation of the clouds $A(\Delta v)$ in {\kms}.
Column 12 lists the number of mean absolute deviations $N_A$ of the
most extreme cloud and column 13 lists a measure of the velocity
asymmetry $\eta$.  See the text for definitions.}

\enddata
\end{deluxetable}

\begin{deluxetable}{clcrrrrrrc}
\tablewidth{0pc}
\tablecaption{\sc Spearman and Kendall Rank Correlation Tests}
\tablehead
{
\multicolumn{2}{c}{Tested Quantities} & &
\multicolumn{3}{c}{Spearman} & 
\multicolumn{3}{c}{Kendall} \\
\cline{1-2} \cline{4-6} \cline{7-9} 
\colhead{Absorption} &
\colhead{Galaxy} &
\colhead{N} &
\colhead{~~~$r_{\rm s}$} & 
\colhead{~~~$P_{\rm s}$} & 
\colhead{~~~$N_{\sigma}$} & 
\colhead{~~~$\tau_{\rm k}$} & 
\colhead{~~~$P_{\rm k}$} & 
\colhead{~~~$N_{\sigma}$} &
\colhead{$P_{\rm MC}$} \\
\colhead{(1)} &
\colhead{(2)} &
\colhead{(3)} &
\colhead{~~~(4)} &
\colhead{~~~(5)} &
\colhead{~~~(6)} &
\colhead{~~~(7)} &
\colhead{~~~(8)} &
\colhead{~~~(9)} &
\colhead{(10)} 
}
\startdata
$W_{\rm r}$     & $B-K$       & 13 & $ -0.54 $ & $ 0.07 $ & $ 1.83 $ & $ -0.43 $ & $ 0.04 $ & $ 2.03 $ & 0.04 \nl 
DR              & $D$         & 14 & $  0.49 $ & $ 0.07 $ & $ 1.84 $ & $  0.40 $ & $ 0.04 $ & $ 2.00 $ & 0.05 \nl 
$N_{\rm c}$     & $D$         & 15 & $ -0.60 $ & $ 0.04 $ & $ 2.02 $ & $ -0.39 $ & $ 0.04 $ & $ 2.01 $ & 0.05 \nl
$N_{A}$         & $L_K$       & 14 & $ -0.58 $ & $ 0.04 $ & $ 2.10 $ & $ -0.43 $ & $ 0.03 $ & $ 2.14 $ & 0.06 \nl 
$A({\Delta v})$ & $D$         & 15 & $  0.47 $ & $ 0.08 $ & $ 1.78 $ & $  0.35 $ & $ 0.07 $ & $ 1.83 $ & 0.06 \nl
$W_{\rm r}$     & $D$         & 13 & $ -0.53 $ & $ 0.06 $ & $ 1.89 $ & $ -0.36 $ & $ 0.07 $ & $ 1.81 $ & 0.08 \nl 
DR              & $B-K$       & 13 & $  0.36 $ & $ 0.18 $ & $ 1.34 $ & $  0.34 $ & $ 0.10 $ & $ 1.63 $ & 0.10 \nl 
$A({\Delta v})$ & $S$         & 13 & $ -0.73 $ & $ 0.01 $ & $ 2.53 $ & $ -0.51 $ & $ 0.01 $ & $ 2.44 $ & 0.14 \nl
\tablecomments
{
Columns 1 and 2 list the absorption and galaxy properties for which
the rank correlation tests were performed. 
Column 3 gives the number of galaxies used in the test.
Columns 4 and 7 tabulate $r_{\rm s}$ and $\tau_{\rm k}$, the Spearman
correlation coefficient and the Kendall $\tau$, respectively.
The probabilities, $P_{\rm s}$ and $P_{\rm k}$, and the number of
standard deviations $N_{\sigma}$ by which the rank coefficients
deviate from their null hypothesis values are listed in columns 
5, 8, 6, and 9, respectively.
In column 10 are the $P_{\rm MC}$ from the Monte--Carlo simulations.
}
\enddata
\end{deluxetable}

\end{document}